\documentstyle[aps,multicol,prl,tighten,graphicx]{revtex}
\begin{document}
\draft
\title{Reducing the Linewidth of an Atom Laser by Feedback}
\author{H.M. Wiseman and L.K. Thomsen}
\date{\today}
\address{School of Science, Griffith University, Nathan, Brisbane,
 Queensland 4111 Australia.}

\maketitle

\begin{abstract}
A continuous atom laser will almost certainly have a
linewidth dominated by the effect of the atomic interaction energy,
which turns fluctuations in the condensate atom
number into fluctuations in the condensate frequency.
These correlated  fluctuations mean that information about the
atom number could be used to reduce the frequency fluctuations,
by controlling a spatially uniform potential. We show that feedback
based on a physically reasonable quantum non-demolition measurement
of the atom number of the condensate {\em in situ}\/ can  reduce
the linewidth enormously.

\end{abstract}

\newcommand{\beq}{\begin{equation}}
\newcommand{\eeq}{\end{equation}}
\newcommand{\bqa}{\begin{eqnarray}}
\newcommand{\eqa}{\end{eqnarray}}
\newcommand{\nn}{\nonumber}
\newcommand{\nl}[1]{\nn \\ && {#1}\,}
\newcommand{\erf}[1]{Eq.~(\ref{#1})}
\newcommand{\dg}{^\dagger}
\newcommand{\rt}[1]{\sqrt{#1}\,}
\newcommand{\smallfrac}[2]{\mbox{$\frac{#1}{#2}$}}
\newcommand{\half}{\smallfrac{1}{2}}
\newcommand{\bra}[1]{\langle{#1}|}
\newcommand{\ket}[1]{|{#1}\rangle}
\newcommand{\ip}[2]{\langle{#1}|{#2}\rangle}
\newcommand{\sch}{Schr\"odinger }
\newcommand{\schs}{Schr\"odinger's }
\newcommand{\hei}{Heisenberg }
\newcommand{\heis}{Heisenberg's }
\newcommand{\bl}{{\bigl(}}
\newcommand{\br}{{\bigr)}}
\newcommand{\ito}{It\^o }
\newcommand{\str}{Stratonovich }
\newcommand{\dbd}[1]{\frac{\partial}{\partial {#1}}}
\newcommand{\sq}[1]{\left[ {#1} \right]}
\newcommand{\cu}[1]{\left\{ {#1} \right\}}
\newcommand{\ro}[1]{\left( {#1} \right)}
\newcommand{\an}[1]{\left\langle{#1}\right\rangle}
\newcommand{\implies}{\Longrightarrow}
\newcommand{\dpar}{\partial}

\begin{multicols}{2}

An atom laser is a continuous source of coherent atom waves, analogous
to an ordinary laser which is a continuous source of
coherent photon waves (light) \cite{fn1,Wis97}. Ideas for creating
an atom laser were published independently by a number of
authors \cite{WisCol95,SprPfaJanWil95,OlsCasDal95,Hol96}, shortly
after the first achievement of Bose-Einstein condensation (BEC) of
gaseous atoms \cite{And95,Bra95,Dav95}.  There have
since been some important advances in the coherent release of
pulses \cite{Mew97,And98} and beams \cite{Hag99,Blo99} of atoms
from a condensate. Since  the condensate is not  replenished in
these experiments, the output coupling cannot continue
indefinitely \cite{Phi00}.
Nevertheless they represent major steps towards achieving a
continuously operating atom laser.

The coherence of an atom laser beam can be defined analogously to that of
an optical laser beam: the atoms should have a relatively small
longitudinal momentum spread, they should ideally be restricted to a
single transverse mode and internal state, and their flux should be
relatively constant \cite{Wis97}. A fourth condition, rarely considered
for optical lasers because it is so easily satisfied, is that the
laser beam be Bose degenerate. This requires that the atom flux be
much larger than the linewidth (the reciprocal of the coherence time)
\cite{Wis97}.
A crucial contributor to the linewidth of an atom laser is the
collisional interaction of atoms, which is negligible for photons.
This is difficult to avoid because it is the collisions between atoms
that enables BEC by evaporative cooling, at present the only method
 for achieving an atom laser.

In this Letter we show that using a feedback mechanism can reduce
the effect of atomic interactions on the atom laser linewidth
by a factor as large as the square root of the atom number.
For a single-mode condensate the dominant
effect of collisions is to turn atom number fluctuations in the condensate
into fluctuations in the energy, which are equivalent to frequency
fluctuations.
By monitoring the
number fluctuations, it is possible using feedback
to largely compensate for the linewidth caused by these frequency fluctuations.
The key practical points are that the measurement does not rely
upon any external condensate phase
reference, and that the control requires only the ability to change
the energy of the atoms, which could be done with a spatially uniform
optical or magnetic field.

We begin by deriving the standard laser linewidth (for non-interacting bosons)
using a simple method which is applied to all later cases. We then derive the
broadened linewidth for an atom laser with strongly interacting atoms.
Finally, we show that feedback based on a  quantum non-demolition (QND)
measurement of atom number in
the the condensate can greatly mitigate this linewidth broadening.

{\em (a) Standard Laser Linewidth.} To derive the standard linewidth we use
the
usual single-mode model of the  laser \cite{Lou73,SarScuLam74}.
Far above threshold, the laser mode
has Poissonian number statistics. In the absence of thermal or other
excess noise, its dynamics are
modeled by the completely positive master equation
\cite{Wis93a,Wis99b}
\beq \label{lme}
\dot{\rho} = \kappa\mu{\cal D}[a\dg]{\cal A}[a\dg]^{-1}\rho
+ \kappa{\cal D}[a]\rho.
\eeq
Here $\kappa$ is the loss rate, $\mu \gg 1$ the stationary mean boson
number, and $a$ the annihilation operator for the laser mode.
The superoperators ${\cal D}$ and ${\cal A}$ are defined as usual:
\bqa
{\cal D}[A]B &\equiv& ABA\dg - {\cal A}[A]B, \\
{\cal A}[A]B &\equiv& (A\dg A B + B A\dg A)/2.
\eqa
That \erf{lme} is of the Lindblad form follows from the identity
\cite{Wis97,Wis93a}
${\cal D}[a\dg]{\cal A}[a\dg]^{-1} = \int_{0}^{\infty} dq{\cal
D}[a\dg e^{-qaa\dg/2}]$.
The stationary solution to this master equation is
\beq
\rho_{\rm ss} = e^{-\mu} \sum_{n} \frac{\mu^{n}}{n!} \ket{n}\bra{n}
     = \frac{1}{2\pi}\int_{0}^{2\pi} d\theta \ket{r
    e^{i\theta}}\bra{r e^{i\theta}}, \label{rhoss}
\eeq
where $r = \sqrt{\mu}$ and $\ket{r e^{i\theta}}$ is a coherent state
\cite{Lou73,SarScuLam74}.

The coherence time of a laser is roughly the time for the phase of the
field to become uncorrelated with its initial value. It is determined
by the stationary first-order coherence function
\beq
g^{(1)}(t) =  {{\rm Tr}[a\dg e^{{\cal L}t}a\rho_{\rm ss}]}/
{{\rm Tr}[\rho_{\rm ss}a\dg a]} \label{cohfun}.
\eeq
A simple and useful definition is  \cite{Wis97}
\beq
\tau_{\rm coh} = \frac{1}{2}\int_{0}^{\infty}|g^{(1)}(t)|dt.
\label{cohtau}
\eeq
Here the $1/2$ is so that, for the standard laser,
$\ell \equiv \tau_{\rm coh}^{-1}$ will be the
standard linewidth. In the cases we consider
it is a very good approximation \cite{fnapprox} to put
$|g^{(1)}(t)| = e^{-i\omega_{0}t}g^{(1)}(t)$
for some frequency $\omega_{0}$.
From \erf{cohfun}, this
allows the integral in \erf{cohtau} to be
evaluated, yielding
\beq
\tau_{\rm coh} \simeq -{{\rm Tr}
[a\dg ({\cal L}-i\omega_{0})^{-1}a\rho_{\rm ss}]}
/{2{\rm Tr}[\rho_{\rm ss}a\dg a]}. \label{numerical}
\eeq

Equation (\ref{numerical}) can be  evaluated numerically, for
example using the {\sc Matlab} quantum optics toolbox \cite{Tan99}.
Analytically, it is easier to use the fact that \erf{cohfun} is
unchanged if $\rho_{\rm ss}$ is replaced by the coherent state
$\ket{r e^{i\theta}}\bra{r e^{i\theta}}$ for arbitrary $\theta$.
Using any suitable phase-space $(\alpha,\alpha^{*})$ representation,
the coherence function then becomes
$
g^{(1)}(t) = \an{\alpha^{*}(t)}/\an{\alpha^{*}(0)}$,
where $\alpha(0)$ has a distribution corresponding to $\ket{r e^{i\theta}}$.
Because fluctuations in the intensity
 $n=|\alpha|^{2}$ are relatively small in a laser with $\mu \gg
1$, the coherence time is well approximated by
\beq
\tau_{\rm coh} \simeq \frac{1}{2}\int_{0}^{\infty} |\langle
e^{-i\delta{\phi}(t)}
\rangle|dt \simeq
\frac{1}{2}\int_{0}^{\infty}e^{-\frac{1}{2}\delta V_{\phi}(t)}dt.
\label{tauphi}
\eeq
 Here
$\delta V_{\phi}(t)$ is the  variance of
$\delta \phi(t) = \arg[\alpha(t)/\alpha(0)]$, and the second
approximation relies on $\delta\phi(t)$ having  Gaussian statistics,
as will be justified below.

For our laser model, the $Q$-function is the most convenient
representation because of the identity
\beq
{\cal D}[a\dg]{\cal A}[a\dg]^{-1}\rho\to
\sum_{k=1}^{\infty}\left( \frac{-\dpar}{\dpar n} \right)^{k}Q(n,\phi),
\eeq
which, since the higher order derivatives are negligible, can
be truncated at $k=2$.
The master equation (\ref{lme}) thus turns into a Fokker-Planck
equation (FPE) for $Q(n,\phi)$ which can be
linearized. Under this FPE,
the number statistics remain those of the initial coherent state,
as does the number-phase covariance $C_{n\phi}$ (i.e. it remains zero),
and the phase has Gaussian statistics with a variances that increases as
$
\delta V_{\phi}(t) = ({\kappa}/{2\mu}) t$.
Substituting this into \erf{tauphi}, we obtain
$
\tau = {2\mu}/{\kappa}$.
This time is precisely the coherence time. Its
reciprocal is the standard laser linewidth $\ell_{0} = \kappa/2\mu$
\cite{Lou73,SarScuLam74,Wis93a,Wis99b}.

{\em (b) Atom Laser Linewidth.} As a model for an atom laser we take the
standard laser master equation (\ref{lme}) and add a term
$
-iC[a\dg a\dg a a ,\rho]
$.
This represents the self-energy of the atoms due to collisions, where
\beq \label{selfenergy}
C = \frac{2\pi\hbar a_{s} }{ m}\int d^{3}{\bf   r}|\psi({\bf
r})|^{4} .
\eeq
Here $\psi({\bf r})$ is the wavefunction for the condensate mode, and $a_{s}$
is the $s$-wave scattering length. For the experiments in Refs.~
\cite{Mew97,And98,Hag99,Blo99}, $\psi({\bf r})$ can be determined
using the Thomas-Fermi approximation, and we use this to obtain the
numerical values which appear later.
The Hamiltonian $Ca\dg a\dg a a$ has no effect on the atom number
statistics. Linearizing the resultant FPE for the $Q$ function
yields the phase-related second-order moments
\bqa
C_{n\phi}(t) &=& \frac{\chi}{2}
(e^{-\kappa t}-1),
\label{covar} \\
\delta V_{\phi}(t) &=& \frac{\chi^{2}}{2\mu}(e^{-\kappa t}+\kappa t - 1) +
\frac{\kappa}{2\mu} t . \label{vphi}
\eqa
Here, we have used $\chi=4\mu C/\kappa$ as a dimensionless parameter for the
strength of the atomic interactions.

This expression for $\delta V_{\phi}(t)$ implies that the integrand
$g^{(1)}(t)$ in
 \erf{tauphi} has the same structure as the analogous expression,
 Eq. (184), derived in
Ref.~\cite{GarZoll98}. This was for a condensate in dynamical equilibrium with
 thermal atoms. Moreover, the three time scales identified in
 Ref.~\cite{GarZoll98} have the same physical origins as those in
 \erf{vphi}.
 The integral in \erf{tauphi}
can be evaluated analytically
in two limits:
\beq
\ell = \tau^{-1} = \left\{ \begin{array}{lll}
    \kappa(1+ \chi^{2})/2\mu & {\rm for} & \chi \ll \sqrt{\mu} \\
    2\kappa \chi /\sqrt{2\pi\mu} & {\rm for} & \chi \gg \sqrt{\mu}
\end{array}  \right. . \label{ellnofb}
\eeq
These correspond to the characteristic time constants of Eqs.
(187) and (186) of Ref.~\cite{GarZoll98} respectively (see
Ref.~\cite{Jak98} for a further discussion).
Our two expressions agree at
$\chi \simeq \sqrt{8\mu/\pi}$, so we have an expression for how $\ell$
scales for all values of $\chi$. The second expression for $\tau$, in the
regime where the nonlinearity is dominant, is familiar as the collapse
time of an initial coherent state in the absence of pumping or damping
\cite{WriWalGar96,ImaLewYou97}.

Using the preliminary atom laser experiments
\cite{Mew97,And98,Hag99,Blo99} as a guide to realistic parameter values, the
dimensionless interaction strength $\chi$ is found to always satisfy
$\chi \gg 1$.
This implies a linewidth for the atom laser
far above the standard limit. If $\chi \agt \mu^{3/2}$, then $\ell$ would be
 larger than the output flux $\kappa\mu$, and the output would
cease to be coherent. It is thus of great interest to find methods for
reducing the linewidth due to atomic interactions.

{\em (c) Effect of QND-based Feedback.}
Atomic interactions do not directly cause phase diffusion. Rather,
they cause a shearing of the field in phase space, with higher
amplitude fields having higher energy and hence rotating faster.
The linewidth-broadening which results is a known effect for optical
lasers with a Kerr ($\chi^{(3)}$) medium \cite{Wat90}.
The shearing of the field
is manifest in the finite value acquired by the covariance
$C_{n\phi}(t)$ in \erf{covar}. It means that information about
the condensate number is also information about the condensate phase.
Hence, we can expect that feedback based on atom number measurements
could enable the phase dynamics to be controlled, and the linewidth
reduced.

It might be thought that one could measure the atom number by
 diverting some of the atom laser output beam. It turns out that
this sort of measurement is effectively useless for reducing the
linewidth. For this reason we consider instead quantum
non-demolition (QND) measurements of atom number, which works very
well \cite{fn2}.

QND atom number measurements
can be performed on the condensate {\em in situ} using dispersive
imaging techniques  \cite{And97}. We consider a far-detuned
probe laser beam of frequency $\omega_p$ and
cross-sectional area $A$ (assumed larger than the condensate) which
passes through the condensate. For simplicity, we will assume that
the distortion of the beam front, and the mean phase shift, are
removed by a suitable ``anti-mean-condensate'' lens.
For a single atom, the interaction Hamiltonian
with the probe beam in the large detuning limit is
\beq \label{deftheta}
\hbar\frac{\Omega_p^{2}}{4\Delta} = \hbar b\dg b \left(
\frac{\hbar\omega_p\Gamma^{2}}{8A\Delta I_{\rm sat}}\right) \equiv
\hbar b\dg b \theta,
\eeq
where $\Omega_{p}$, $\Delta$, $\Gamma$ and $I_{\rm sat}$ have their usual
meaning. Here $b$ is the annihilation operator for the probe beam,
normalized so that $\hbar \omega_p b\dg b$ is the beam power.
The effective interaction Hamiltonian for the whole condensate,
minus the mean phase shift, is thus
\beq \label{Hint}
H_{\rm int} = \hbar \theta (a\dg a-\mu) b\dg b,
\eeq
where $\theta$, defined in \erf{deftheta}, is the
phase shift of the probe field due to a single atom.

From \erf{Hint}, the back-action of the probe
fluctuations on the condensate can be evaluated using the techniques
of Ref.~\cite{Wis94a}, and results in the extra phase diffusion
\beq
\dot{\rho} = \beta^{2}{\cal D}[e^{-i\theta(a\dg a-\mu)}]
\rho \simeq N {\cal D}[a\dg a]\rho.
\eeq
Here the approximation requires $\sqrt{\mu}\theta \ll 1$, and
\beq \label{defparamD}
N = \beta^{2}\theta^{2}=P\theta^{2}/\hbar\omega_p,
\eeq
where $\beta=\an{b_{\rm in}}$ and $P$ is the mean beam power.
Equation (\ref{Hint}) also gives the output probe field \cite{Wis94a}
\beq
b_{\rm out} = e^{-i\theta (a\dg a-\mu)}b_{\rm in} \simeq
b_{\rm in} - i\theta(a\dg a-\mu)\beta,
\eeq
where the same approximation has been used. The condensate
number fluctuations can thus be measured by homodyne detection of the
output phase
quadrature
\beq
I^{\rm Y}_{\rm out} = -ib_{\rm out}+ib_{\rm
out}\dg \simeq I^{\rm Y}_{\rm in} -2\beta\theta(a\dg a - \mu).
\eeq

In order to control the phase dynamics of the condensate, we wish to use
the homodyne current to modulate the phase. This can
be done, for example, by applying a uniform magnetic field or far-detuned
light field across the whole condensate. We model this by the Hamiltonian
\beq
H_{\rm fb}(t) = -\hbar a\dg a \sqrt{\frac{\kappa\lambda}{4\mu}}
\int_{0}^{\infty}h(s)I_{\rm out}^{\rm Y}(t-s)ds,
\eeq
where $h(s)$ is the response function of the feedback loop and $\lambda$
is a dimensionless measure of the feedback strength.
Neither the measurement nor the feedback affect the
atom number statistics.
In the ideal limit of instantaneous feedback,
$h(s) \to \delta(s)$ and we can apply the Markovian theory of
Ref.~\cite{Wis94a}.
The total master equation for the atom laser is then
\bqa
\dot\rho &=& \kappa\mu{\cal D}[a\dg]{\cal A}[a\dg]^{-1}\rho
+ \kappa{\cal D}[a]\rho -iC[a\dg a\dg aa,\rho]
\nl{+} iN\sqrt\frac{\lambda}{\nu} [a\dg a\dg a a,\rho]
+ N\left(1+\frac{\lambda}{\eta\nu}\right){\cal D}[a\dg a]\rho.
\label{mefb}
\eqa
Here we have allowed for a detection efficiency $\eta$ \cite{Wis94a},
and dropped terms corresponding to a frequency shift,
and defined a dimensionless parameter $\nu = 4\mu N/\kappa$.

Proceeding as before, we find the $Q$ phase variance:
\bqa
\delta V_{\phi}(t) &=& \frac{1}{2\mu}\left(\chi-\sqrt{\nu\lambda}\right)^{2}
( e^{-\kappa t} + \kappa t - 1) \nl{+}
 \frac{1}{4\mu}\left(2+\nu+\frac{\lambda}{\eta}\right) \kappa t.
\eqa
If the feedback is to reduce the linewidth, we need
$(\chi-\sqrt{\lambda\nu})^{2} \ll 4\mu$ and $\eta$
must be large enough that $\eta\mu \gg 1$.
In this case, \erf{tauphi} has a simple analytical solution:
\beq
\ell=\tau^{-1} = \frac{\kappa}{4\mu}\left[2+\nu+\frac{\lambda}{\eta}
+2\left(\chi-\sqrt{\nu\lambda}\right)^{2}
\right].
\eeq

Assume $\chi$ is large enough that $2\sqrt{\eta}\chi> 1$, which
is the physically interesting regime where the self-energy is important. Then
$\ell$ is minimized for a feedback strength of
$\lambda=\sqrt{\eta}\chi-1/2$, and a measurement strength of
$\nu = \lambda/\eta$. The optimum linewidth in this case is
\beq
\ell_{\rm min} =
\frac{\kappa}{2\mu} \ro{1+\frac{\chi}{\sqrt{\eta}}-\frac{1}{4\eta}}.
\label{ellfb}
\eeq
Note that for a large interaction strength $\chi$ there is an optimum
measurement strength $\nu \simeq \chi/\sqrt{\eta}$
(or equivalently $N \simeq C/\sqrt{\eta}$)
independent of the output coupling rate $\kappa$. From \erf{mefb} we
see that this effectively cancels
the self-energy term in the master equation (since $\lambda
= \eta \nu$). A weak measurement will
give poor information about the atom number, with a high
noise-to-signal ratio. Feeding back the noisy current to counter the
shearing caused by the nonlinearity will thus add large
phase fluctuations to the condensate.
On the other hand, if the measurement is
too strong the measurement back action in the form of phase diffusion, as
discussed above \erf{defparamD}, will itself dominate the linewidth.

In Fig.~1 we plot the approximate  analytical expressions for the linewidth in
the absence [\erf{ellnofb}] and presence [\erf{ellfb}] of
feedback as a function of
nonlinearity $\chi$. We also plot exact numerical results using
\erf{numerical}, for $\mu= 60$. The agreement is
very good. It is evident that the QND feedback offers
a linewidth
much smaller than that without feedback for all values of $\chi >
1/2\sqrt{\eta}$. In fact for $\chi \gg 1$ the reduction factor approaches a
maximum of $\sqrt{8\mu/\pi}$. Most importantly, with feedback, the laser
output remains coherent until $\chi \agt \mu^{2}$, a much higher value than
the
$\mu^{3/2}$ which applies in the absence of feedback.

\begin{figure}
\includegraphics[width=0.45\textwidth]{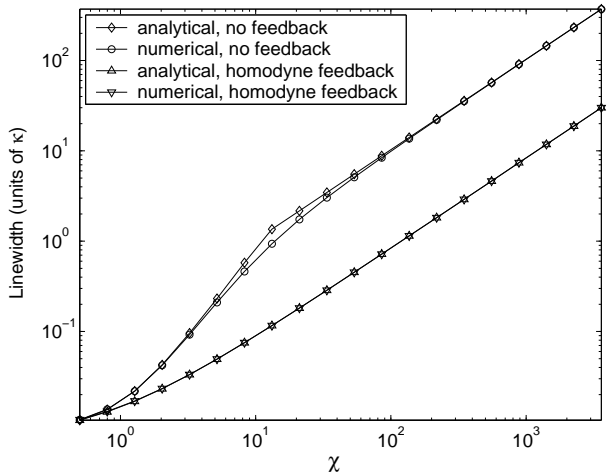}
\vspace{0.2cm}
\caption{\narrowtext Atom laser linewidth for $\kappa=\eta=1$ and
$\mu=60$,
plotted with and without feedback using both analytical and numerical
methods (see text).}
\protect\label{linewidth}
\end{figure}

Let us summarize. An atom laser will almost certainly suffer great linewidth
broadening due to the collisional self-energy of the atoms. This is
because the self-energy produces a correlation between the number and
phase fluctuations of the condensate. We have
shown that this broadening can be enormously reduced by controlling the
phase of the condensate
based on a QND measurement of the number of atoms in the condensate.
The factor of reduction can be as large as the square root of
the number of atoms in the condensate (that is, a factor of perhaps
$10^{3}$).

A question of interest is, how easy is it to obtain a QND measurement
of sufficient strength to optimize the feedback? We have seen above
that we require $\nu \simeq \chi$, which is equivalent to $N \simeq C$.
For the typical BEC parameters of
Refs.~\cite{Mew97,And98,Hag99,Blo99}, $C \sim 10^{-2}$s$^{-1}$.
It may be verified from \erf{defparamD} that it is very easy to
obtain a measurement strength
$N$ this large, even with $\Delta \gg \Gamma$.

A related question is,
how much of a problem is atom loss due to
spontaneous emission by atoms excited by the detuned probe beam? The
rate of this loss (ignoring reabsorption) is
$\Gamma \times (\textrm{Excited Population})$. We would like the ratio of this
 loss rate to  the
output loss rate $\kappa \mu$ to be small.
In the $\Delta \gg \Gamma$ limit, this ratio is given by
\beq \label{ratio}
\Gamma\frac{(\Gamma/2\Delta)^{2}(I/2I_{\rm sat})\mu}{\kappa \mu}
 \sim  \frac{2\chi I_{\rm sat} A}{\hbar\omega_p\Gamma\mu}.
\eeq
For physically reasonable parameters of $\chi \sim 10^{3}$,
$I_{\rm sat} \sim 10 $W$/$m$^{2}$, $A \sim 10^{-11}$m$^{2}$,
$\omega_p \sim 3\times 10^{15}$s$^{-1}$, $\Gamma \sim 3\times
10^{6}$s$^{-1}$, and
$\mu \sim 10^{6}$, \erf{ratio} is indeed small ($\sim 10^{-1}$).

In conclusion, there appear to be no fundamental reasons
that this proposal could not be put
to good use when a continuous atom laser is realized.

{\flushleft \em Acknowledgments.}
HMW is deeply indebted to W.D. Phillips for the idea of controlling
atom laser phase fluctuations using atom number measurements,
and for subsequent insightful comments on this
work.

\end{multicols}


\begin{references}
\bibitem{fn1}
Of course, some optical lasers have a pulsed rather than continuous output.
However, as argued in Ref.~\cite{Wis97}, in order for the term ``atom
laser'' to mean something more than merely an atomic Bose-Einstein condensate
released from its trap, it is useful to impose the condition that a
true atom laser have a continuous output.
\bibitem{fnapprox}
We have shown analytically and numerically that the
relative error caused by this approximation is $O(\mu^{-1})$.
\bibitem{fn2}
 This is perhaps not surprising, since
feedback based on demolition photodetection cannot produce squeezed
states, whereas that based on QND detection can \cite{WisMil94a}
\end{references}
\end{document}